# Power-to-Frequency Conversion in Cryogenic Sapphire Resonators

Eugene N. Ivanov and Michael E. Tobar

*Abstract* - We studied how the cryogenic sapphire resonator responds to fast variations of the dissipated microwave power. The experiments were carried out with sapphire resonators cooled to 6 K at frequencies around 11 GHz. We found that the power-to-frequency conversion of the resonator depends on Fourier frequency as the transfer function of the 1st-order low-pass filter with corner frequency close to the resonator's loaded half-bandwidth. Having measured the power-to-frequency conversion of the cryogenic sapphire resonator, we predicted the phase noise of the microwave oscillator based on such resonator.

## I. INTRODUCTION

The Cryogenic Sapphire Loaded Cavity (SLC) resonators are crucial elements of secondary frequency standards or "sapphire clock" used in frequency metrology [1, 2], high-precision tests of fundamental physics [3-4], and radio-astronomy [5]. In [6], cryogenic SLC resonators were proposed to generate low-noise microwave signals for applications in Doppler radars.

One of the noise mechanisms limiting the spectral purity of the microwave signals generated with the cryogenic SLC resonators is the power-to-frequency conversion in the resonator. This effect is associated with the radiation pressure-induced deformation of the sapphire crystal and power-induced fluctuations of its temperature [7]. In the experiments with "sapphire clocks," Chang *et al*. measured the dependence of the cryogenic SLC resonator frequency on dissipated microwave power [8]. The measurements were performed in a steady-state regime long after the resonator's temperature control system had "absorbed" the stepwise power change of the incident microwave signal. That work left unanswered the question about the dynamic response of the cryogenic SLC resonator to rapid variations of microwave power. Answering this question is essential for predicting the noise performance of cryogenic microwave oscillators.

In this correspondence, we (i) describe an experimental technique for measuring the power-to-frequency conversion of the cryogenic resonator as a function of Fourier frequency and (ii), based on the obtained results, evaluate the phase fluctuations of the cryogenic sapphire oscillator.

## II. MEASUREMENT APPARATUS

Fig. 1 shows a schematic diagram of an apparatus for measuring the dynamic response of a cryogenic resonator to power fluctuations. The setup features a microwave frequency synthesizer, the resonator itself, a mixer, and a few auxiliary components which form a Frequency Discriminator (FD).

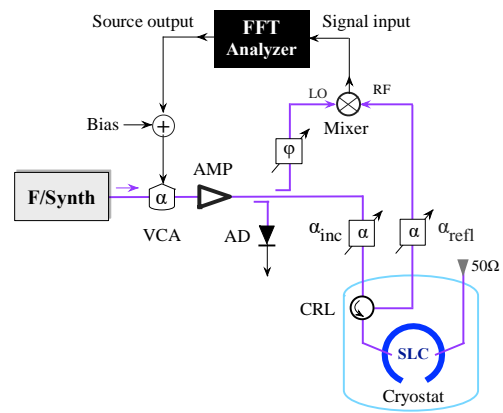

Fig. 1. Experimental setup: VCA – voltage-controlled attenuator, CRL – circulator, AMP – power amplifier, AD – amplitude detector, F/Synth – microwave frequency synthesizer, and SLC – sapphire loaded cavity resonator (kept at temperature close to 5 K with a pulse-tube cryocooler).

We tuned the FD by modulating the synthesizer frequency and varying the phase difference between two signals interacting at the mixer to maximize the amplitude of the FD output AC signal. To generate an AM-modulated signal for resonator interrogation, we used the Voltage Controlled Attenuator (VCA) driven by the voltage noise from the FFT analyser (see Fig. 1). Such modulation technique proved superior to the direct AM-modulation of the frequency synthesizer, as it was accompanied by more than an order of magnitude less spurious modulation of phase.

We placed two mechanical attenuators, $\alpha_{inc}$ and $\alpha_{refl}$, in the path of the incident and reflected signals, respectively. When measuring the power-to-frequency conversion at different levels of incident power $P_{inc}$, we varied $\alpha_{inc}$ while keeping the total attenuation $\alpha_{inc}\alpha_{refl}$ constant. This approach eliminated the need to calibrate the measurement system every time the incident power changed. Also, it was easier to differentiate the effect of the power-to-frequency conversion (proportional to the $P_{inc}$) from the spurious influences caused by imperfections

[1] Eugene Ivanov, Quantum Technologies and Dark Matter Labs, Department of Physics, The University of Western Australia, 35 Stirling Highway, Crawley, 6009 (eugene.ivanov@uwa.edu.au)

This work is supported by the Australian Research Council (grants CE170100009 and CE200100008)

Michael Tobar, Quantum Technologies and Dark Matter Labs, Department of Physics, The University of Western Australia, 35 Stirling Highway, Crawley, 6009 (michael.tobar@uwa.edu.au)

of the VCA and residual amplitude sensitivity of the FD (which were independent of the $P_{inc}$).

## III. SAPPHIRE RESONATORS

The experiments were carried out with three almost identical Sapphire Loaded Cavity (SLC) resonators. Each resonator featured a copper shield (with a diameter of 86…88 mm and a height of 51…53 mm) housing a HEMEX sapphire spindle with a diameter of 51 mm and a height of 30 mm. According to the sapphire spindle's manufacturers, its rotational and crystal axes were aligned within a few degrees. Each resonator was excited in the WGH-modes using magnetic loop probes [9]. The size of the copper shield and probe's position relative to the sapphire spindle were adjusted to maximize the contrast of the resonator's transmission peak. After each adjustment, the resonator was cooled close to 6 K to measure its S-parameters.

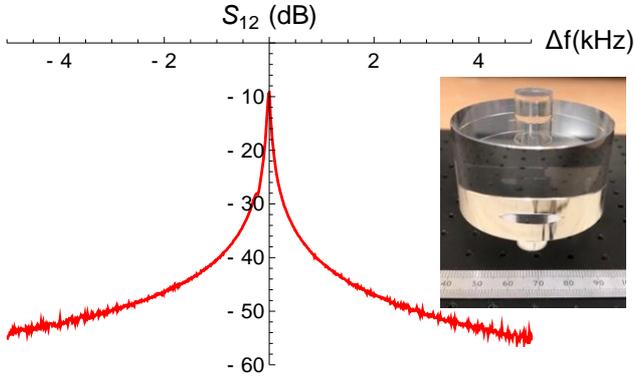

Fig. 2. Amplitude transmission coefficient of the cryogenic SLC resonator vs offset from resonance. The inset shows the sapphire spindle

Fig. 2 shows the transmission coefficient of one of the SLC resonators as a function of offset frequency from the resonance ($f_{res} \sim 11.2$ GHz). The shape of the transmission peak is close to Lorentzian with a 3dB bandwidth of 52 Hz. The peak contrast, i.e., the separation between its top and the pedestal (located 70…90 kHz away from the peak's top), was approximately 65 dB. All three cryogenic resonators exhibited the transmission peaks similar that shown in Fig. 2. Each resonator was mode-matched to the incoming signal: at least 95% of the incident power was absorbed in the tuned resonator.

## IV. NOISE ANALYSIS AND DISCUSSION

Below, we present a list of equations we used to infer the resonator power-to-frequency conversion from the results of voltage noise measurements. We begin with an analytical expression for the total voltage noise at the mixer output:

$$\delta u_\Sigma = \sqrt{\delta u_{synth}^2 + \delta u_{induced}^2} \qquad (1)$$

where the first term under the square root reflects the synthesizer's phase noise contribution, whereas the second term gives a combined effect of three different noise mechanisms caused by the voltage noise applied to the VCA:

$$\delta u_{induced} = S_{FD}\, \delta f_{res} + S_{FD}\, \delta\phi_{vca}\, \mathcal{F} + \mathcal{R}_{FD}\, \delta m \qquad (2)$$

where $\delta f_{res}$ is the resonant frequency fluctuations due to its sensitivity to power, the $\delta\phi_{vca}$ is the phase fluctuations due to VCA voltage-to-phase conversion, $\delta m$ is the AM-index fluctuations of the incident signal. The last term in (2) results from the residual AM-sensitivity of the frequency discriminator.

The analytical expressions for the fluctuating terms in (2) are following:

$$\delta f_{res} = \frac{df_{res}}{dP_{diss}}\, 2\, P_{diss}\, \delta m \qquad (3.1)$$

$$\delta m = \frac{1}{2\alpha}\frac{d\alpha}{du}\, \delta u_{noise} \qquad (3.2)$$

$$\delta\phi_{vca} = \frac{d\phi_{VCA}}{du}\, \delta u_{noise} \qquad (3.3)$$

where $df_{res}/dP_{diss}$ is the resonator power-to-frequency conversion to be measured, $P_{diss}$ is the power dissipated in the resonator, $\alpha$ is the VCA insertion loss at a given bias voltage, $d\phi_{vca}/du$ is the VCA voltage-to-phase conversion, and $\delta u_{noise}$ is the voltage noise applied to the VCA. Coefficients $S_{FD}$ and $\mathcal{R}_{FD}$ in (2) are given by

$$S_{FD} = \chi\, \sqrt{P_{inc}}\, \frac{2\beta}{1+\beta}\, \frac{1}{\Delta f_L + i\mathcal{F}}\, \sin(\Phi) \qquad (4.1)$$

$$\mathcal{R}_{FD} = \chi\, \sqrt{P_{inc}}\, \frac{2\beta}{1+\beta}\, \frac{i\mathcal{F}}{\Delta f_L + i\mathcal{F}}\, \cos(\Phi) \qquad (4.2)$$

where $\chi$ is the mixer power-to-voltage conversion, $P_{inc}$ is the power incident on the resonator, $\beta$ is the resonator coupling coefficient, $\Delta f_L$ is the resonator loaded half-bandwidth, $\mathcal{F}$ is the Fourier frequency, and $\Phi$ is the phase difference between two mixer ports.

As follows from (1-4), to improve the confidence with which the resonator frequency fluctuations are detected, one must: (i) use the low-phase noise frequency synthesizer; (ii) chose VCA with the lowest level of voltage-to-phase conversion, and (iii) minimize the FD spurious AM-sensitivity $\mathcal{R}_{FD}$.

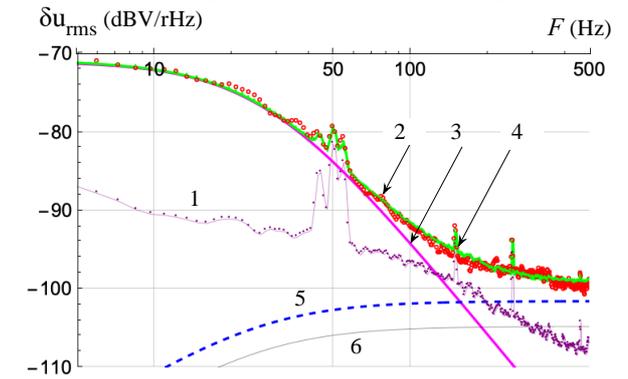

Fig. 3. Voltage noise spectra at the output of the Frequency Discriminator due to: (1) Synthesizer phase fluctuations. (2) Combined effect of all noise sources (experiment). (3) Resonator power-to-frequency conversion (4) Combined effect of all noise sources (calculations) (5) VCA spurious phase modulation. (6) FD residual amplitude sensitivity.

We used Eqns. 1-4 to infer the power-to-frequency conversion $df_{res}/dP_{diss}$ from the voltage noise spectra at the FD output. The noise measurements were carried out under the following conditions: (i) The microwave power dissipated in the resonator $P_{diss}$ was close to 18 dBm. (ii) The FD sensitivity to slow frequency modulation ($F_{mod} \ll \Delta f_L$) was approximately 16500 mV/kHz. (iii) The induced amplitude noise of the incident signal $\delta m$ (measured with the amplitude detector AD in Fig. 1) was approximately $10^{-3}\ 1/\sqrt{Hz}$.

First, we measured FD voltage noise without any modulation applied to the VCA (trace 1 in Fig. 3). Next, we modulated VCA bias with white voltage noise to produce an AM-signal for interrogation of the cryogenic resonator. The voltage noise spectrum measured in such a case is given by trace 2 in Fig. 3.

Trace 3 in Fig. 3 is the voltage noise spectrum associated with the resonator power-to-frequency conversion, i.e., it corresponds to the first term in (2). It was computed for the power-to-frequency conversion given by

$$\frac{df_{res}}{dP_{inc}} = \frac{A}{1 + i\,\mathcal{F}/\Delta f_L} \qquad (5)$$

where $A \sim 0.12$ Hz/mW and $\Delta f_L = 26$ Hz.

Eq. 5 describes the transmission coefficient of the 1st order low-pass filter with the corner frequency close to the resonator's loaded half-bandwidth.

The fit (5) was valid for all three SLC resonators tested. Thus, for the second SLC resonator, we measured $\Delta f_L = 23$ Hz and $A \sim 0.14$ Hz/mW. In both cases, no dependence of the parameter A on dissipated microwave power $P_{diss}$ (varying from 3 to 20 dBm) was found. For the third resonator, the DC power-to-frequency conversion was a strong function of dissipated power. For example, the inferred value of A was close to 0.3 Hz/mW at $P_{diss} = 18$ dBm. Decreasing the $P_{diss}$ to 3 dBm increased the conversion efficiency to 2 Hz/mW.

The non-linearity of the third SLC resonator was attributed to the relatively high concentration of paramagnetic $Cr^{3+}$ ions in sapphire [10,11]. We attempted to reduce the power-to-frequency conversion by saturating the $Cr^{3+}$ Electron Spin Resonance (ESR). For that, we pumped one of the SLC "auxiliary" modes in the vicinity of the $Cr^{3+}$ ESR ($f_{ESR} \sim 11.445$ GHz). This proved impractical; to observe an appreciable reduction in power-to-frequency conversion we had to inject almost 10 mW into the "auxiliary" mode.

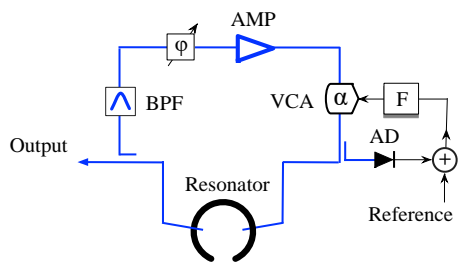

Fig. 4. Proposed cryogenic microwave loop oscillator

Knowing the frequency-power sensitivity of the cryogenic sapphire resonator, one can estimate the phase noise of the microwave oscillator based on such a resonator (see Fig. 4). Thus, Fig. 5 shows the phase noise spectra of the 11.2 GHz microwave loop oscillator calculated at $2\Delta f_L = 50$ Hz and the power of the output signal $P_{out} = 1$ mW. Trace 1 in Fig. 5 corresponds to the oscillator based on the conventional microwave amplifier. The noise spectrum comprises the inclined part and the horizontal pedestal; the former is due to the 1/f - phase noise of the microwave amplifier, and the latter is due to thermal noise [12]. Trace 2 shows the phase noise spectrum of the oscillator based on the reduced noise amplifier [13]. In such a case, the 1/f – noise is significantly reduced due to the use of an ultra-sensitive interferometric phase detector as a sensor of a high-gain noise suppression system. Finally, trace 3 shows the phase noise spectrum caused by resonator power-to-frequency conversion. It was computed at $A = 0.12$ Hz/mW, assuming that oscillator power was stabilized with the control system based on the double-balanced mixer acting as an amplitude detector. The average power dissipated in the cryogenic resonator was assumed to be 20 mW. This was the maximum power at which the resonator-induced phase noise did not exceed the limit set by the electronics of the loop oscillator.

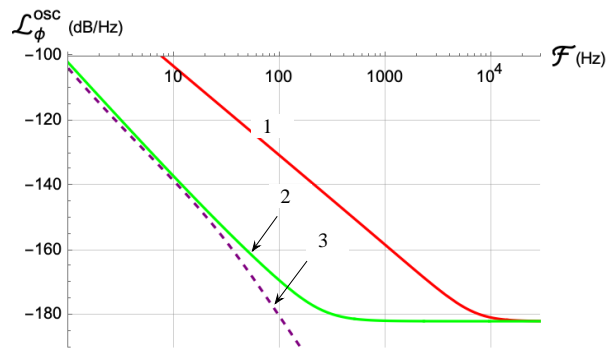

Fig. 5. Predicted phase noise spectra at 11 GHz. (1) Conventional loop oscillator. (2) Loop oscillator based on reduced noise amplifier. (3) Limit due to power-to-frequency conversion in the cryogenic sapphire resonator at A= 0.12 Hz/mW and $P_{diss} = 20$ mW.

One can compare the phase noise performance of the cryogenic sapphire oscillator to that of an optical frequency synthesizer based on femtosecond laser technology [14,15]. Such a comparison shows that the phase noise spectra of the two signal sources are similar at Fourier frequencies below 30 Hz. Yet, at $\mathcal{F} > 100\ Hz$, the phase noise of the cryogenic oscillator is almost two orders of magnitude lower than that of the photonic system. We expect an additional 5…7 dB improvement in oscillator phase noise relative to that in Fig.5 by exciting the SLC resonator in a mode near 10 GHz, i.e., by widening the gap between the SLC operating frequency and the Cr+3 ESR.

V. CONCLUSION

We measured the power-to-frequency conversion of the cryogenic sapphire resonators. Three sapphire resonators of the same geometry and size were tested at temperatures near 6 K. It was found that power-to-frequency conversion varies with Fourier frequency as the transmission coefficient of the 1st-order low-pass filter with the corner frequency close to the resonator loaded half-bandwidth. Also, we found that the static value of the power-to-frequency conversion strongly depended on the concentration of paramagnetic Cr3+ ions. Last but not least, our simulations showed that the resonators with a low concentration of Cr3+ ions were well suited for developing microwave oscillators with very low phase and amplitude fluctuations.

## REFERENCES

[1] J. Guéna, M. Abgrall, D. Rovera et al, "Progress in atomic fountains at LNE-SYRTE," *IEEE Trans. on UFFC* (2012) 59:391–409. doi:10.1109/tuffc.2012.2208.

[2] S. Weyers, V. Gerginov, M. Kazda, "Advances in the accuracy, stability, and reliability of the PTB primary fountain clocks," *Metrologia* (2018) 55:789–805.

[3] P. Stanwix, M. Tobar, P. Wolf et al, "Test of Lorentz Invariance in Electrodynamics Using Rotating Cryogenic Sapphire Microwave Resonators," Phys. Rev. Letters, vol. 95,040404(2005)

[4] H. Muller, P. Stanwix, M. Tobar et al, "Tests of Relativity by Complementary Rotating Michelson-Morley Experiments," *Phys. Rev. Letters*, vol. 99, 050401 (2007).

[5] V. Giordano, S. Grop, P. Bourgeois et al, "New-generation of cryogenic sapphire microwave oscillators for space, metrology and scientific applications," Review of Scientific Instruments 83, 085113 (2012), DOI: 10.1063/1.4747456.

[6] E. Ivanov and M. Tobar, "Noise Suppression with Cryogenic Resonators," Microwave and Wireless Components Letters, vol. 31, Issue 4, pp. 405-408, April 2021, DOI: 10.1109/LMWC.2021.3059291, Print ISSN: 1531-1309, Online ISSN: 1558-1764.

[7] V. Braginsky, V. Mitrafanov, and V. Panov, "Systems with Small Dissipation*"* (University of Chicago, Chicago, 1985), p. 78.

[8] S. Chang, A. G. Mann, A. N. Luiten, and D. G. Blair, "Measurements of Radiation Pressure Effect in Cryogenic Sapphire Dielectric Resonators," v. 79, no. 11, pp. 2141-2145, 1997.

[9] C. Locke, E. Ivanov, J. Hartnett et al, "Invited Article: Design techniques and noise properties of ultra-stable cryogenically cooled sapphire-dielectric resonator oscillators", Review of Scientific Instruments, v. 79, no. 5, pp. 051301 1-12, 2008.

[10] W. Farr, D. Creedon, M. Goryachev et al," Ultrasensitive microwave spectroscopy of paramagnetic impurities in sapphire crystals at millikelvin temperatures," Physical Review B 88, 224426 (2013).

[11] V. Giordano, S. Grop, P.-Y. Bourgeois, Y. Kersale, and E. Rubiola," Influence of the electron spin resonance saturation on the power sensitivity of cryogenic sapphire resonators," J. Appl Phys, 116, 054901 (2014).

[12] E. Ivanov and M. Tobar, "Real time noise measurement system with sensitivity exceeding the standard thermal noise limit," IEEE Trans. On UFFC, vol. 49, no. 8, 2002, pp. 1160-1161.

[13] E. Ivanov, M. Tobar and R. Woode, "Microwave interferometry: application to precision measurements and noise reduction techniques," IEEE Trans. on UFFC, vol. 45, no. 6, Nov. 1998, pp. 1526-1537.

[14] T. Fortier, A. Rolland, F. Quinlan et al, "Optically referenced broadband electronic synthesizer with 15 digits of resolution", *Laser Photonics Review*, 1–11 (2016) / DOI 10.1002/lpor.201500307.

[15] X. Xie, R. Bouchand, D. Nicolody et al, "Photonic microwave signals with zepto-second-level absolute timing noise", *Nature Photonics*, v. 11, January 2017, pp. 44-47.